%% file: main.tex
\documentclass[manuscript]{acmart}

\usepackage{booktabs}
\usepackage{tabularx}
\usepackage{multirow}
\usepackage{calc}
\usepackage[capitalise]{cleveref}

\AtBeginDocument{%
  \providecommand\BibTeX{{%
    \normalfont B\kern-0.5em{\scshape i\kern-0.25em b}\kern-0.8em\TeX}}}

\setcopyright{acmcopyright}
\copyrightyear{2021}
\acmYear{2021}
\setcopyright{acmlicensed}\acmConference[C\&C '21]{Creativity and Cognition}{June 22--23, 2021}{Virtual Event, Italy}
\acmBooktitle{Creativity and Cognition (C\&C '21), June 22--23, 2021, Virtual Event, Italy}
\acmPrice{15.00}
\acmDOI{10.1145/3450741.3465253}
\acmISBN{978-1-4503-8376-9/21/06}

\newcommand{\system}{CharacterChat}
\newcommand{\conceptBotShaping}{Progressive Manifestation}

\newcommand{\modeGuided}{Guided Conversation Mode}
\newcommand{\modeOpen}{Open Conversation Mode}

\newcommand{\pct}[1]{#1\,\%}

\newcommand{\lastaccessed}{\textit{last accessed 28.01.2021.}}

\newcommand{\studyOneN}{7}
\newcommand{\studyTwoN}{8}

\makeatletter
\newcommand\footnoteref[1]{\protected@xdef\@thefnmark{\ref{#1}}\@footnotemark}
\makeatother

\definecolor{rowcol}{rgb}{0.9,0.9,0.9}

\definecolor{DanielsColor}{rgb}{0.9,0.6,0.1}

\definecolor{OliversColor}{rgb}{0,0.3,0.9}

\definecolor{highlightboxColour}{rgb}{0.95,0.95,0.95}

\newcommand{\highlightbox}[2]{
    \vspace{\baselineskip}
    \noindent
	\colorbox{highlightboxColour}{
	\begin{minipage}{.96\linewidth}
	    $\rightarrow$\quad \textbf{Design Takeaway #1:}
	    #2 
	\end{minipage}
	}
	\vspace{0.5em}
}

\begin{document}

\title{\textit{\system:} Supporting the Creation of Fictional Characters through Conversation and \conceptBotShaping{} with a Chatbot}

\renewcommand{\shorttitle}{\textit{\system:} Supporting the Creation of Fictional Characters with a Chatbot}

\author{Oliver Schmitt}
\affiliation{%
  \institution{Research Group HCI + AI, Department of Computer Science, University of Bayreuth}
  \city{Bayreuth}
  \country{Germany}
}
\email{oliver.schmitt.de@outlook.de}

\author{Daniel Buschek}
\orcid{0000-0002-0013-715X}
\affiliation{%
  \institution{Research Group HCI + AI, Department of Computer Science, University of Bayreuth}
  \city{Bayreuth}
  \country{Germany}
}
\email{daniel.buschek@uni-bayreuth.de}

\renewcommand{\shortauthors}{Schmitt and Buschek}

\begin{abstract}
\input{sections/abstract}

\end{abstract}



\maketitle

\input{sections/introduction}

\input{sections/related_work_and_concept_dev}

\input{sections/implementation}

\input{sections/user_studies}

\input{sections/discussion}

\input{sections/conclusion}

\begin{acks}
We thank the writers for participating in our study, and Christina Schneegass and Sarah Theres Völkel for their feedback on the manuscript.
This project is funded by the Bavarian State Ministry of Science and the Arts and coordinated by the Bavarian Research Institute for Digital Transformation (bidt).
\end{acks}

\bibliographystyle{ACM-Reference-Format}
\bibliography{bibliography}

\end{document}

%% file: sections/abstract.tex
We present \textit{\system}, a concept and chatbot to support writers in creating fictional characters. 
Concretely, writers progressively turn the bot into their imagined character through conversation.
We iteratively developed \textit{\system} in a user-centred approach, starting with a survey on character creation with writers (N=30), followed by two qualitative user studies (N=\studyOneN{} and N=\studyTwoN). 
Our prototype combines two modes: (1) \textit{Guided prompts} help writers define character attributes (e.g. \textit{User: ``Your name is Jane.''}), including suggestions for attributes (e.g. \textit{Bot: ``What is my main motivation?''}) and values, realised as a rule-based system with a concept network. (2) \textit{Open conversation} with the chatbot helps writers explore their character and get inspiration, realised with a language model that takes into account the defined character attributes. 
Our user studies reveal benefits particularly for early stages of character creation, and challenges due to limited conversational capabilities. We conclude with lessons learned and ideas for future work.

%% file: sections/introduction.tex
\section{Introduction}

\begin{figure*}[!t]
\centering
\includegraphics[width=\textwidth]{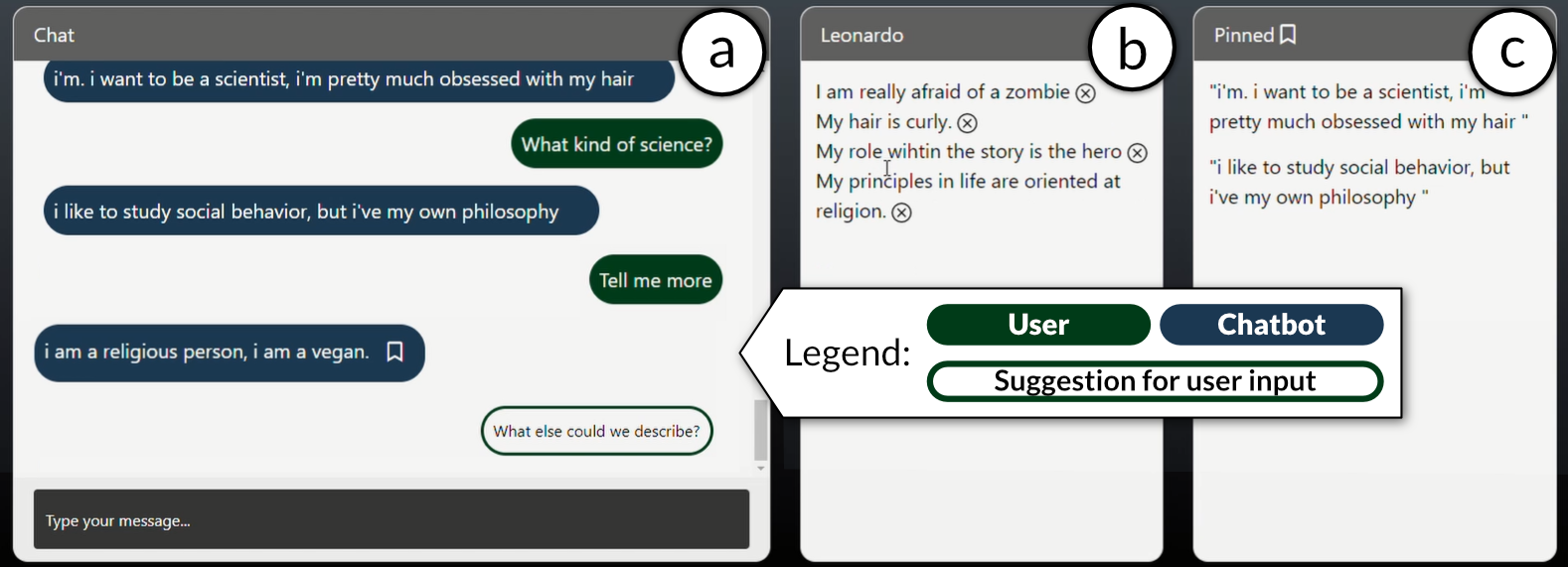}
\caption{Our UI: (a) \textit{chat} between user (right bubbles, green) and bot (left bubbles, blue); (b) \textit{character view}, shows attributes defined so far via the chat (can be deleted via ``x'' buttons, added after study 1); (c) \textit{pinboard view} with lines generated by the bot's open dialogue model and pinned by the user in the chat, e.g. for inspiration and later use. Legend added for the figure, not part of the UI.} 
\label{fig:ui}
\end{figure*}

\begin{figure*}[!t]
\centering
\includegraphics[width=\textwidth]{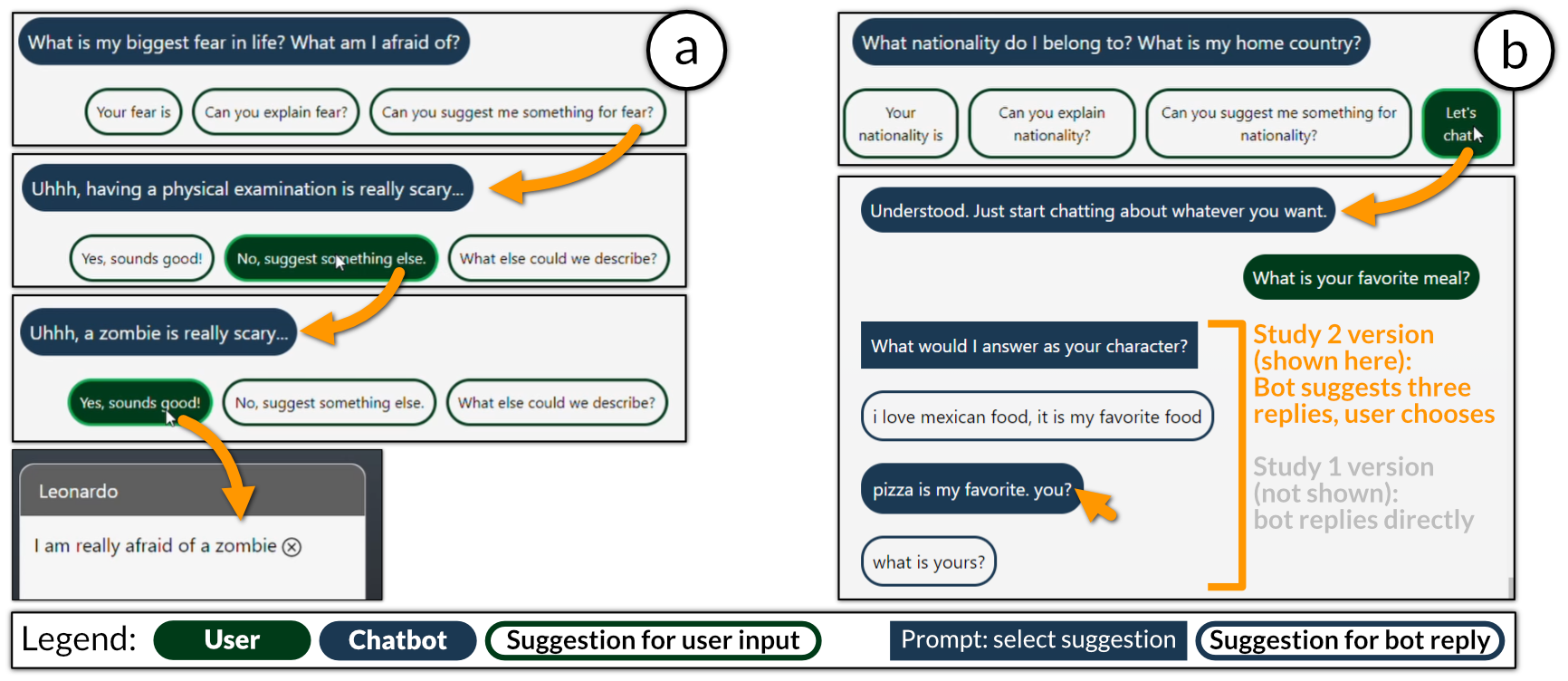}
\caption{Using \textit{\system}'s two modes: (a) In \textit{\modeGuided}, the bot suggests attributes to define (here: ``biggest fear''), randomly drawn from a list informed by the literature and our survey. As shown, the user can then ask for suggestions for concrete values (generated with a concept network~\cite{Speer2017conceptnet}), and accept or reject them. Here, the user accepts ''zombie'', which is then listed in the character view. The user could have freely entered anything instead. (b) Saying ``Let's chat'' transitions to \textit{\modeOpen}, in which the bot responds to anything (here: user asks about a favourite meal). The figure shows the improved version (i.e. after study 1), in which the bots suggests three potential replies (here: user selects the 2nd one, ``pizza''), instead of just giving one reply directly.} 
\label{fig:modes}
\end{figure*}

Fictional characters are at the heart of creative narrative work today: 
This includes novels, movies, TV series, stage plays, computer games, and (pen and paper) role playing games.
Creating compelling characters thus presents an important and impactful activity for writers in such projects.
This is a difficult task because compelling characters require depth, complexity or ``roundness''~\cite{Fishelov1990, Forster1927, Hochman1985, Rimmon2003}.
This challenge has motivated a variety of writing guides and techniques for fictional characters (e.g. see~\cite{Clark2008, Egri1972, Kirschenbaum2016}) and their integration into a story (e.g. Campbell's \textit{Monomyth} template~\cite{Campbell2008hero}).
Digital tools for characters and story integration exist as well (e.g. \textit{Dramatica}, \textit{The Novel Factory}, \textit{Papyrus Autor}\footnote{\label{footnote:papyrus}\url{https://dramatica.com/}, \url{https://www.novel-software.com/}, \url{https://www.papyrus.de/}, \lastaccessed}). These tools focus on bookkeeping of existing characters (i.e. tracking character info, scenes), yet not ideation or ``prototyping'' of new characters, and do not utilise the computational medium to bring characters ``alive''. %

To address this lack of creativity support for character creation%
, we explore the new idea of a \textit{conversational} tool approach in this context. This idea is motivated by two aspects: (1) In Human-Computer Interaction (HCI), conversation is a \textit{user interface paradigm} (e.g. chatbots), recently also successfully employed to interact with text~\cite{terHoeve2020}. (2) In writing, conversations are an \textit{expressive stage for characters}, often used to reveal a character's personality to the reader (e.g. via dialogue or internal monologue). Thus, we investigate how we might utilise this ``match'' (i.e. conversation as both a UI paradigm and creative expression), with the following research question:
\textit{How might a conversation-based tool -- a chatbot -- support writers in creating fictional characters?}

To address this question, we propose a chatbot concept called \textit{\conceptBotShaping}: Writers turn the bot into a fictional character by chatting with it.
Our prototype, \textit{\system} (\cref{fig:ui}), realises this with two modes (\cref{fig:modes}): 

\begin{itemize}
    \item \textit{\modeGuided:} For defining the character, the chatbot offers \textit{rule-based prompts} (e.g. \textit{Bot: ``What is my main motivation?''}), informed by research from Media Science and Narratology, and our survey. The bot can further suggest concrete attributes values (e.g. hair colors), based on a \textit{concept network} (\textit{ConceptNet}~\cite{Speer2017conceptnet}).
    \item \textit{\modeOpen:} For exploring the character, a \textit{Deep Learning dialogue model} enables open conversations with the writer. This model takes into account the character information elicited through the prompts so far. %
\end{itemize}

The writer may interweave guided and open conversation with the chatbot, switching between these two modes.
As a result, writers can use the bot both to define the character (e.g. \textit{User: ``Your name is Jane.''}) and to get inspiration (e.g. \textit{User: ``What is your hair colour?'', ``What have you done last week?''}).

We iteratively developed \textit{\system} in a user-centred approach:
First, we conducted an online survey with 30 writers to gain insight into their process of character creation and inform our design. 
Second, we implemented our tool and evaluated its use for character creation in a qualitative user study (N=\studyOneN), leading to an improved UI and chatbot.
Finally, we evaluated this improved version in a separate qualitative study (N=\studyTwoN).

In summary, we contribute:
(1) \textit{\system}, a chatbot that supports writers in creating fictional characters via our concept of \textit{\conceptBotShaping}; and (2) insights from its user-centred development and evaluation with writers. %

%% file: sections/related_work_and_concept_dev.tex
\section{Informing and Developing the \textit{\system} Concept}
Here we report on background and steps of our concept development, based on the literature and an online survey.
Conceptually, we explored the initial idea of a conversation-based tool, towards an informed first design, by combining character models and creation techniques in Media Science and Narratology, with creativity support tools and conversational UIs in HCI. For brevity, we first describe this literature directly together with a report of its relation to our design choices (\cref{sec:concept_theory}). In our process, these choices were also informed by the survey (\cref{sec:survey}) and we reflect on its value in this regard in our discussion (\cref{sec:discussion_of_methods}).

\subsection{Theoretical Background and Related Work}\label{sec:concept_theory}

\subsubsection{Creating Characters in Storytelling}\label{sec:concept_characters_creativity}
The etymology of \textit{character} points to a stamping tool -- figuratively, a person's stamp of personality~\cite{Eder2010}. Our \textit{\conceptBotShaping} concept picks up on this with the idea of painting a ``blank'' chatbot into a character by adding information.
Therefore, we build on the model of
\citet{Forster1927} who distinguishes characters as \textit{``flat''} (based on a single quality) or \textit{``round''} (more than one factor). Related considerations of complexity appear in more extensive later models (e.g. \citet{Hochman1985}, \citet{Fishelov1990}). While both types can be useful, we focus on round characters because they present the harder challenge: \citet{Forster1927} emphasises achieving roundness by considering \textit{multiple} qualities, which motivates our \textit{\modeGuided}. Here, our bot prompts writers to define many attributes, and can suggest concrete attribute values for inspiration on demand with a concept graph~\cite{Speer2017conceptnet}. 
Practically, for the construction of fictional characters we consider writing guides and frameworks (e.g. \citet{Egri1972}, \citet{Mckee1999}, \citet{Seger1994}), which all support writers with conceptual categories: Concretely, we build on the attributes from the character ``bone structure'' by \citet{Egri1972}, which covers the three categories physiology (e.g. age, eye and hair colour), psychology (e.g. values, ambitions, skills), and sociology (e.g. social background, education, community), enriched with attributes elicited in our survey (see \cref{sec:survey_descriptor_categories}).

A related character model by \citet{Rimmon2003}\footnote{Often attributed to be based on work by Joseph Ewan.} articulates \textit{complexity}, \textit{development}, and \textit{penetration into inner life}. These motivate our \textit{\modeOpen}: It attempts to technically enable the character to gain complexity by revealing its ``inner life'' and development in the chat. For instance, the writer could ask the character-bot about its past actions and memories to dig deeper and get inspiration. Moreover, while our tool does not explicitly support story plotting, asking the character-bot about goals and opinions (cf. a character's ``dramatic goal'' and ``position'' by \citet{Field1984}) might also stimulate plot ideas by the writer.
Finally, both modes together can be compared to a distinction made by \citet{Mckee1999} about ``characterization'' (observable attributes -- included in our guided prompts) vs ``true character'' (below the surface, including decisions -- e.g. asking the bot about decisions and actions in the open chat).

\subsubsection{Creativity Support Tools for Writing}
\textit{\system} is a concept and tool in the tradition of creativity support tools (CSTs)~\cite{Shneiderman2002, Shneiderman2007}. %
Based on the definition by \citet{Frich2019}, CSTs are ``employed to positively influence users of varying expertise in one or more distinct phases of the creative process''. 
In this light, we focus on the phase of character creation. %
Specifically, expressed in the framework by \citet{Shneiderman2002}, our tool addresses three of the tasks listed therein: (1)~\textit{Thinking} -- conversing with the character to foster associations and ideas. (2)~\textit{Exploring} -- our chatbot is a ``simulation model'' of the character, explorable via conversation, in particular in \textit{\modeOpen}. (3)~\textit{Composing} -- our chatbot supports ``step-by-step'' composition of the character, in particular in \textit{\modeGuided}.

Examples of commercial tools for creative writing include \textit{Dramatica}, \textit{The Novel Factory} and \textit{Papyrus Autor}\footnoteref{footnote:papyrus}, which support tracking characters, relations, appearances in scenes, etc.
While these tools thus broadly support writing, we focus on \textit{creating} characters and explore representations beyond \textit{passive} character sheets: We instead ``manifest'' the character in a new \textit{active} manner -- a chatbot -- to foster inspiration and bringing characters ``alive''. 

Writing tools in the (HCI) literature support academic writing~\cite{Strobl2019}, stories~\cite{Clark2018, Watson2019storyprint}, poetry~\cite{Hafez2017}, metaphors~\cite{Gero2019chi}, and slogans~\cite{Clark2018}, using NLP~\cite{Clark2018, Gero2019chi, Hafez2017, Strobl2019}, visualisations~\cite{Watson2019storyprint} or images~\cite{Chang2011}: For instance, \textit{Storyprint} plots scenes and characters on a ``circular time axis''~\cite{Watson2019storyprint}. %
NLP researchers explored (controlled) story plot generation with AI~\cite{Tambwekar2019}, including (reinforcement learning) agents~\cite{Aylett2013, Martin2018, Riedl2004}, which (inter)act as characters to create the plot. In contrast, our agent (chatbot) is not meant to generate stories automatically but to support human writers in creating characters. Others used agents as a proxy audience during story creation~\cite{ONeill2009}. Further work focused on \textit{author} characteristics: \textit{InkWell}~\cite{Gabriel2015} uses computational optimisation to generate text variants based on data-driven author personas. In summary, in contrast to previous work, %
we explore the use of an AI agent (chatbot) for ideation, and as a character simulation, to support writers when creating characters. %

Regarding industry and the arts, \textit{BanterBot} by Google's Creative Lab\footnote{\url{https://experiments.withgoogle.com/banter-bot}, \textit{last accessed 27.01.2021}} allows writers to describe a character in a few sentences to prime a chatbot. A promotional video %
reports inspiring writers yet no scientific study was conducted, the tool is not accessible, and it is unclear how its design was informed. This motivates our user-centred design and evaluation, and sharing our prototype. Conceptually, we go beyond \textit{BanterBot} in both UI/interaction design (more than a chat view, \cref{fig:ui}) and creativity support (integration of a concept graph, multiple suggestions, the two modes, \cref{fig:modes}).

\subsection{Empirical Background: Initial User Research}\label{sec:survey}
We further informed our design with an online survey with (amateur) writers: After giving informed consent, people were asked about writing background and approaches, and creating characters. Concretely, we elicited further character attributes beyond those recommended in the literature (\cref{sec:concept_characters_creativity}), and gained insights into writing processes.

\subsubsection{Participants}
We recruited 30 participants (17 female, 13 male) via an online forum for writers. Their mean age was 33 years (range: 14 to 63). 
With multiple selections allowed, 26 people (\pct{86.7}) stated to write as a hobby, for six people (\pct{20}) it was (also) a part-time job, and for one person a full-time job. Two people (additionally) listed other aspects.
Moreover, 21 people (\pct{70}) stated to write for themselves, 17 people (\pct{56.7}) wrote (also) for exchange with others, and 18 (\pct{60}) wrote for publication. Three people (additionally) listed other aspects.

\subsubsection{Results}\label{sec:survey_descriptor_categories}

\paragraph{General Writing Approach}
With multiple selections allowed, 26 people (\pct{86.7}) said to typically begin a writing project with a rough idea; eleven (\pct{36.7}) start without planning; seven (\pct{23.3}) start with a character; also seven (\pct{23.3}) start with the setting; five (\pct{16.7}) start with a plot outline; four (\pct{13.3}) mentioned other approaches, such as defining a central conflict.
On average, people gave two answers; even for a single author approaches can vary.

\paragraph{Creating and Developing Characters}
Asked as open questions, \textit{inspiration} for characters included music, existing characters, internet research, books and movies, and real life people.
Used \textit{tools and templates} included writing character arcs, and creating mind maps, biographies and family trees.
Reported \textit{issues} point towards undesirably stereotypical creations: Characters turn out to be too close to employed templates or means of inspiration, such as too closely resembling an archetype or famous person. They can also lack distinctiveness, or turn out to be too flat or shallow, matching well the considerations in the literature (\cref{sec:concept_characters_creativity}).

\paragraph{Describing Characters}
We asked what kind of information is important to describe characters. People could list as many aspects as they liked. To give an overview here, as grouped by one of the researchers and checked by the other, people mentioned aspects relating to social life (18 unique descriptors), appearance (16), motivation/goals (16), personality traits (15), quirks (15), background/story (12), and further personal data (12), thoughts (8), interests/hobbies (7), moral values (7), likes/dislikes (7), and story function (7). %

\subsubsection{Discussion \& Implications}
In short, we found great diversity in writing approaches, in line with related work~\cite{Kirschenbaum2016}. %
As a key takeaway, writers overall cover a wide range of aspects when describing their characters, yet not everyone thinks about everything. Combined with the writers' reported issues on flat characters, and the literature (\cref{sec:concept_characters_creativity}), this indicates an opportunity to support writers in exploring character attributes to facilitate creating round, complex characters.
The results further provided a concrete list of such attributes to inform our implementation. %

%% file: sections/implementation.tex
\section{Implementation}\label{sec:implementation}
This section describes our prototype \textit{\system} system. It was iteratively improved based on the results of the first user study (see \cref{sec:improvements_after_study_1}). To avoid redundancy, here we directly describe the improved (final) version.

\subsection{User Interface and Interactions}
\cref{fig:ui} shows the user interface of our prototype, implemented as a web app. It has three main parts: a chat view for the conversation, a character view that shows the attributes defined so far, and a pinboard view that shows dialogue pieces that the user has ``bookmarked'' in the chat view. Readers are also invited to view the video.\footnote{See project repository: \url{https://osf.io/dyatj/}}
Moreover, \cref{fig:modes} shows and explains two example interaction flows for the two modes.

\subsection{Backend and Chatbot}

Technically, the implementation of our chatbot combines three modules, explained next.

\subsubsection{Understanding Intents}
We used \textit{DialogFlow}\footnote{\url{https://cloud.google.com/dialogflow}, \lastaccessed} to recognise users' intents from their chat input: In particular, this module handles the intents involved in the guided parts (see below), plus switching between guided and open modes.

\subsubsection{Guided Conversation with a Rule-based System and Concept Graph}\label{sec:implementation_guided}
We implemented the \textit{\modeGuided} with a simple rule-based approach which draws a yet undefined attribute at random to suggest to the user (e.g. the ``biggest fear'' in \cref{fig:modes}a). The list of 31 attributes is based on the survey and literature~\cite{Egri1972} and could be flexibly extended in the future. For each attribute, we also wrote a short explanation that the bot can provide upon request (see middle suggestion in \cref{fig:modes}a top box).
In addition, we integrate \textit{ConceptNet}~\cite{Speer2017conceptnet}\footnote{\url{https://conceptnet.io/}, \lastaccessed} as a knowledge base to suggest attribute values (e.g. the ``physical examination'' and ``zombie'' in \cref{fig:modes}a). Concretely, we use \textit{ConceptNet's} relations (mainly ``is-a''), such as ``A is-a C'' or ``B is-a C'', to (randomly) suggest ``A'' or ``B'' as possible values for an attribute ``C''.

\subsubsection{Open Conversation with a Neural Language Model}
For the \textit{\modeOpen}, we used a publicly available Deep Learning language model released by \textit{HuggingFace}~\cite{Wolf2019}, described as ``a conversational AI with a persona''\footnote{\url{https://medium.com/huggingface/how-to-build-a-state-of-the-art-conversational-ai-with-transfer-learning-2d818ac26313}, \lastaccessed}: It generates text based on both the chat history and a few sentences about the chatbot's ``personality''. We refer the reader to the cited references for technical details. Conceptually, with this model, we condition text generation both on the conversation in the chat view (\cref{fig:ui}a) and the attribute statements in the character view (\cref{fig:ui}b). In this way, we technically realise our concept of ``manifesting'' the fictional character in the chatbot as the conversation progresses.

%% file: sections/user_studies.tex
\section{User Study I: First Prototype}

We conducted a user study to evaluate our \textit{\system} concept and prototype and get first feedback from writers on the design choices and usefulness of the tool.

\subsection{Study Design and Apparatus}
The study was approved by our institution. Due to the \textit{covid19} pandemic, it was a remote study: People used our prototype web-app via their browsers, while sharing their screen in a video call with a researcher. 
We focused on qualitative insights, and thus favoured detail over sample size, via observation, think-aloud, and interviews, complemented with chat log analysis and questionnaires. %

\subsection{Participants and Procedure}
We recruited seven amateur writers (6 female, 1 prefer not to disclose) via university networks. Their mean age was 25 years (range: 24-27). All stated to write as a hobby, while one also wrote as a part-time job. Motivations included writing for themselves (all people), for exchange with others (4 people), for publication (1), and for pen \& paper games (1). 

With recruitment, we asked people to think of a writing project for which they would like to develop a character, to increase our task's relevance. All but one person prepared such a context.
After informed consent and a questionnaire (demographics, writing background), %
we introduced our prototype. People had 20 minutes to create a character. A researcher was in the video call for questions and observations, encouraging ``thinking aloud''.
Semi-structured interviews then asked about inspiration, usability, conversation with the bot, and further use and ideas. %
Finally, people filled in the \textit{Creativity Support Index} (CSI) questionnaire~\cite{Cherry2014}. These sessions took about 40 minutes. People were paid 10\,€.

\subsection{Results}

\subsubsection{Interviews \& Feedback During Use}

We took notes during the sessions and interviews and further analysed them by watching the recordings. We report feedback per interview question, plus related comments during use. \cref{tab:studies_example_characters} (top two rows) shows example characters created in the study.

\paragraph{Inspiration}
When asked about getting ideas for the character, everyone replied positively:
Except for one, everyone mentioned the guided questions (e.g. discovering aspects such as ``religion'' or ``fear''), that helped to ``make the [character] a little deeper'' (P2).
Regarding ideas that came from the chatbot's generated text, interviewees mentioned, for example, ``the death of the character through a car accident'' (P4), ``that the character owns a cat'' (P7), or statements that seem to reveal personality, such as the bot saying ``I'm not afraid of flying. I'm afraid of how people treat me'' (P5).

\paragraph{Feedback on Specific Features}
All writers except one mentioned the guided conversation, when asked to reflect on important and helpful features:
The guided mode helped to consider new aspects -- ``sometimes you don't even think about [the aspect] and then you remember oh yes this could be quite interesting'' (P5). %
One person explicitly preferred the open conversation (``can manage to define the [attributes] on my own'', P7).
Four out of seven also mentioned the open conversation as one of the most important features, mainly due to the possibility ``to communicate more'' (P7).
Having both modes in the tool was also explicitly mentioned as helpful -- for example, P6 highlighted ``the combination [...] because
one would be difficult without the other in practice''. 
Finally, the suggestion of attribute values was also mentioned as helpful, yet less prominently so, and P4 reported that ``they were not really fitting''.

\paragraph{Conversation with the Chatbot}
Overall, the chatbot did not feel like an authentic conversation: The main reason was the lack of context within the chabot's responses. For example, P7 said that ``it is funny but not always that fitting''. In addition, four found that the personality of the character was not sufficiently reflected in the statements: Especially, ``if you have developed the character further, it might be counterproductive, because the chat is not as advanced as you would like it to be'' (P2). Still, five people could have imagined having longer conversations with the chatbot, including for entertainment and curiosity about the possible conversation topics.

\paragraph{Deeper Understanding of the Character}
We asked if people had a deeper understanding of their character after using the tool. Opinions were divided:
Some stated their image of the character was not influenced. Especially those who had already a fixed idea felt less influenced. 
In contrast, others found that the conversation underlined the character (e.g. P3 got a ``better feeling [for the character] through chat history''), or added certain aspects, such as a ``love interest'' (P6). Similarly, P2 said they reconsidered their stereotypical heroine because the bot said ``I've never killed anyone''. %

\paragraph{Further Use of the Created Character}
Writers would not use their study's character in their project right away, partly due to the limited time spent on it in the study. While two people said they ``can imagine that [further elaboration of the character] would be possible'' in the tool (P3), three others saw its focus ``to give input at the beginning'' (P4). 
Especially main characters lacked detail. Two people could imagine using the tool, for example, ``if you have a secondary character [...] and want to give [it] a little more depth'' (P5).

\paragraph{Further Ideas and Use of the Tool}
Everyone could imagine including the tool in their process -- in particular for creating new characters, but also for further description of partly defined characters. As P7 put it: ``there are many characterisation lists where you also go through these questions in detail [...] but I think it is a bit more natural with the bot.''
P4 also thought about advantages for people without experience in character creation, such as new players of a pen and paper scenario. %
Writer's block and lacking motivation were mentioned as further potential uses of the tool.

Further ideas for the tool included a progress indicator for the character, attribute visualizations, relationships between multiple characters, and support for inputting a (partly defined) character. In summary, five people especially desired more interaction possibilities, such as, for instance, editing and removing defined attributes.

\begin{table*}[!t]
\centering
\footnotesize
\renewcommand{\arraystretch}{1.25}
\begin{tabularx}{\textwidth}{@{}p{4em}XX@{}}
\toprule
\textbf{Character} & \textbf{Attributes} (from guided mode, here shortened for presentation) & \textbf{Pinned Statements} (from bot in open mode, pinned by the user) \\
\midrule
``Lucrezius'' (study 1) & 32 years; IQ 140; hobby: listening to Jazz; strange quirk: loving cabbage; principles: theory; attitude: single minded, racist; high social class; family dead; religion ``?''; work: scientist. & ``Thank you! I am a scientist.'', ``I like Jazz.'', ``I'm quite the celebrity.'', ``I'm in the lab and I'm actually studying how to make cabbage.'', ``I'll die in the car. How about you?'', ``I have a PhD in genetics.''\\
``Miri'' (study 1) & goal: new experiences; hobbies: playing tennis, having parties; attitude: living life to the fullest, not caring a lot about others, pursuing my own goals; fear: afraid of ending up like my parents; upper middle class; born in Innsbruck. & ``I'm not rich, I'm just into music.'', ``I love animals. I like to play with my cat.'' \\
``Amy'' (study 2) & violet eyes; dark purple hair; female; 19 years; abdominous; from Mediterranean Sea; a mermaid; born in the bottom of the ocean; really good at typing; astute; first thing you see is that I'm positive; hobby: karaoke. & ``I like all kinds of horror movies.'', ``No, but I do have a very large tattoo on my back.'', ``It is a tattoo of a dolphin.'', ``I had to get a tattoo when I was younger.''\\
``Asimir'' (study 2) & from Galifarn; goal: to help my village regain independence; strange quirk: magic ability and connection to the god Khelemvor; diathesis; principles: a conviction; metallic eye color; hair: flock. & ``I'm a little bit of an introvert, but I try to make things work.'', ``I love them, they're so supportive.'', ``I did, I'm not sure why, but I want to help my people.'', ``It is all good, but it is hard to find friends in the world.'', ``Her name is Abigail. She's a great person.'', ``I love cats. They're so cute... I've an affinity for them.''\\
\bottomrule
\end{tabularx}
\caption{Four examples of characters created in the studies with attributes defined in \textit{\modeGuided} (cf. \cref{fig:ui}b) and statements from \textit{\modeOpen} pinned by the user (cf. \cref{fig:ui}c). Attributes shortened for presentation (e.g. Amy: ``The color of my hair is dark purple'' $\rightarrow$ ``dark purple hair''), text lightly edited (user typos/grammar, capitalisation in bot statements).}
\label{tab:studies_example_characters}
\end{table*}

\subsubsection{Creativity Support Index}

The prototype overall achieved a CSI score of 67.78 out of 100 points. Table~\ref{tab:studies_csi_results} shows the full results, computed and presented as described in the related work~\cite{Cherry2014}. %
The average factor scores describe how the users rated the prototype on that aspect (scale) with a maximum possible value of 20.
Complementary, with the average factor counts, the CSI questionnaire assesses the users' relative importance of these aspects, with a maximum value of five. 
Thus, the results indicate strong \textit{Enjoyment}, and good \textit{Results worth effort} and \textit{Exploration}. Considering importances (i.e. factor scores), participants valued \textit{Exploration} and \textit{Expressiveness} most highly. %
\textit{Immersion} had an intermediate rating. %
People rated the prototype very low on \textit{Collaboration}.  %
Note that the CSI questionnaire here asks questions about collaboration with ``other people''~\cite{Cherry2014}, which was not a function in this prototype. The very low factor score indicates that people also did not find this aspect to be relevant here. We decided to still keep this question for our study to not change the original CSI questionnaire and computation.

\begin{table*}[!t]
\centering
\small
\newcolumntype{R}{>{\centering\let\newline\\\arraybackslash\hspace{0pt}}m{5em}}
\begin{tabular}{@{}p{8em}RRcRRcRR@{}}
\toprule
\multirow{2}{*}{\textbf{Scale}} & \multicolumn{2}{c}{\textbf{Avg. Factor Score (FS)} (max 20)} && \multicolumn{2}{c}{\textbf{Avg. Factor Count} (max 5)} && \multicolumn{2}{c}{\textbf{Avg. Weighted FS} (max 100)} \\
& Study 1 & Study 2 && Study 1 & Study 2 && Study 1 & Study 2\\
\midrule
Results worth effort & 15.86 & 17.38 && 1.86 & 2.00 && 29.45 & 34.75 \\
Exploration          & 15.71 & 16.63 && 4.43 & 4.38 && 69.59 & 72.73 \\
Collaboration        & ~3.14 & 10.50 && 0.57 & 0.50 && ~1.78 & ~5.25 \\
Immersion            & 10.28 & 11.38 && 2.57 & 2.50 && 26.45 & 28.44 \\
Expressiveness       & 11.71 & 14.63 && 3.86 & 3.50 && 45.18 & 51.19 \\
Enjoyment            & 18.00 & 18.13 && 1.71 & 2.13 && 30.86 & 38.52 \\ 
\midrule
\multicolumn{2}{l}{\textbf{Creativity Support Index (CSI)}} & && & && \textbf{67.78} & \textbf{76.96} \\
\bottomrule
\end{tabular}
\caption{Results of the Creativity Support Index (CSI) questionnaire~\cite{Cherry2014} from the first and second study: Each factor (column 1) is assessed via two 10-point Likert items, summed to \textit{factor scores} (column 2). Comparative questions assess the factors' relative importances, which result in the \textit{factor counts} (column 2). Scores and counts are combined to yield \textit{weighted factor scores} (column 3) and an overall \textit{CSI score} (bottom row).}
\label{tab:studies_csi_results}
\end{table*}

\subsubsection{Chat Logs}
The mean dialogue length was 64 lines (SD: 21).
We analysed the logs for (technical) problems and found issues on \textit{robustness} (e.g. handling incorrectly spelled user input), \textit{bugs} (e.g. last attribute defined again with next input), %
and \textit{concepts} (e.g. wrong suggestions for hair style).
Regarding interaction, the start was not always clear (e.g. just ``Hi'' works but defining the name directly such as ''Hi Jane'' was not supported), and some \textit{confusing mode switches} (e.g. bot responds by talking about its family in open mode while user wanted to define ``family'' as an attribute).
While we focused on issues here, the logs also showed that the chat system had generally worked as intended.

\subsection{Prototype Improvements Based on the Study}\label{sec:improvements_after_study_1}
Based on these results we decided to make the following improvements to our prototype for the next iteration:

\subsubsection{Giving Users More Choice} 
We changed the system such that the UI shows three suggested replies in \textit{\modeOpen}, instead of giving just one reply directly (\cref{fig:modes}b). %
Only the user's chosen reply is added to the chat history. This addresses limitations of the bot, in that users are not stuck with an odd reply but may ignore it in favour of another. Moreover, revealing more options might facilitate inspiration and exploration.

\subsubsection{Giving Users More Control} 
We added a ``delete'' function to the list of the collected character attributes (small ``X'' next to each attribute, see \cref{fig:ui}). This allows users to easily remove attributes, which is intended to improve robustness for users and facilitate exploration that involves changing one's mind about an attribute.

\subsubsection{Making Mode Switches More Explicit} 
We adjusted the prototype such that switching between the two modes is more explicit. This makes intent recognition more robust and avoids confusion on the users' part: Concretely, instead of switching on the fly based on intent recognition, the improved chatbot displays a suggested user input to \textit{``Let's chat''} (see \cref{fig:modes}b) after the user has defined three attributes in \textit{\modeGuided}. Conversely, in \textit{\modeOpen} it shows a suggested user input to switch back (\textit{``What else could we describe?''}).

\subsubsection{Further Improvements} 
Finally, we fixed bugs and made smaller adjustments to the UI and backend and logic based on the study insights (e.g. revised UI design, and accounting for further variations of user input).

\section{User Study II: Improved Prototype}

We conducted another study to evaluate our improved prototype (\cref{sec:improvements_after_study_1}). Study design and procedure were identical to the first study.
However, we gave people 10 minutes more to interact with the bot, based on the feedback of study~1.
We recruited eight amateur writers (4 female, 4 male) via university networks, with a mean age of 26 years (range: 21-34). One had also participated in the first study. All but one stated to write as a hobby. Two were (also) part-time authors. Motivations included writing for themselves (5 people), for exchange with others (6) and for publication (3). People were paid 10\,€.
To avoid redundancy, the following report focuses on insights into the impact of the changes, and comparisons to the first study. \cref{tab:studies_example_characters} (bottom two rows) shows example characters created in the study.

\subsection{Interviews \& Feedback During Use}

\subsubsection{Inspiration}
Four people gave clear positive feedback on inspiration, as the system asked for attributes they ``hadn't thought about [...] at all before'' (P2) and suggested values ``that were very helpful'' (P6). 
Three people stated to get ``not completely new ideas but [...] a new approach'' (P4).
The person that had also participated in study 1 found that it did not help with ideas ``this time'' (P1), because for their character in study 2 they had a more specific image in mind beforehand. 
Nevertheless, all interviewees positively mentioned getting ideas from the bot, such as ``being cynical'' (P6), ``having a dolphin tattoo on the back'' (P8) or that ``it’s all good but it’s hard to find friends in the world'' (P6).

\subsubsection{Feedback on Specific Features}
Similar to study 1,  three people considered especially the guided conversation as helpful for defining aspects they ``would not have thought about otherwise'' (P6) and that it ``felt like a red thread at the beginning'' (P5). 
The open conversation was rated positive by three people, for example, due to ``the fact that it's still expanding on things based on what it's been given [and that] it's just fascinating to talk to it'' (P4).
The suggestion of attribute values was especially important for two people, including for new reasons beyond study 1: ``author bias is eliminated or reduced [... to avoid characters that] can do everything and are super great, which is unrealistic [and] no one wants to read something like that'' (P2).

\subsubsection{Conversation with the Chatbot}

Feedback on the conversation was more positive than in study 1:
Three people explicitly stated that they felt like chatting with the character. For instance, P4 found ``it's like you're talking directly to the character and not to some language program'' and that it ``doesn't feel like [the bot] is just repeating stuff, but making something out of [the attributes]''.
However, two criticized that the conversation did not go smoothly, and three encountered answers contrary to defined attributes. %
P1 found the bot's personality was not reflected in its dialogue's style: %
``if [... a] character is 14, [... the bot] doesn't speak like 14 [year olds] and there are [...] typical phrases missing''. P7 also missed this. 
Related, P5 saw the system ``more as [...] an interactive notepad, but not necessarily as a conversation partner'' (P5).
Still, in contrast to study 1, all people said they could imagine having longer chats with the bot.

\subsubsection{Deeper Understanding and Further Use of the Character}
In contrast to study 1, six people gave positive feedback on a deeper understanding, and some critical views related more to the study (P3: too little time) than the system. Four writers stated that their image or aspects of their prepared character had changed due to the interaction with the chatbot.
In contrast to more hesitant outlooks in study 1, five people could imagine including the created character in their stories, or, if there already was a rough design, to ``take [aspects] on board'' (P4).
The other three people stated that some things were missing, such as typical utterances, or that it ``develops anyway when writing [the story]'' (P7).

\subsubsection{Further Ideas and Use of the Tool}
As in study 1, all people could imagine further using the tool in their writing process, although P2 stated that it would be necessary to have a save/load function to directly ``ask the [existing] character for his opinion''. Similar to impressions in study 1, the tool would be suitable for the ''exploratory initial phase'' (P4) of character development or ``for the supporting characters'' (P8). %
Especially for later writing stages, the system ``would need improvements'' (P5). Four people asked us to contact them if the tool was available online.

\subsection{Creativity Support Index \& Chat Logs}
The prototype scored 76.96 CSI points, indicating improvement over the first one (67.78), although we run no statistical tests due to the small sample. Nevertheless, descriptively, ratings improved for all aspects (\cref{tab:studies_csi_results}). 
The mean dialogue length was 80 lines (SD: 44).
The chat logs confirmed our bug fixes and that the new features worked (e.g. mode switches, choosing among generated text options). %

%% file: sections/discussion.tex
\section{Discussion and Takeaways}
Here we reflect our design choices, methods and results and highlight lessons learned that we expect to be useful more broadly for designing systems that use (co-creative) AI and/or conversational UIs to support creative writers.

\subsection{\textit{\system} Supports Early Stages of Character Creation}
As an overall result, the concept was accepted by writers. With respect to the user needs addressed by the design, the user studies indicate that interacting with the chatbot is particularly interesting in the \textit{early} stage of creating a character. Here, the conversation helps to capture fundamental character aspects, and to get ideas for further aspects to consider. It can also help enrich secondary characters. %
Moreover, capabilities of current chatbots are less restricting in early stages where even an ``odd'' reply by the bot might not contradict existing ideas of the character, or might even have inspirational value. This was also revealed by the person who participated in both studies -- once with an early character and once with a more developed one.

\subsection{Mode Switches in Creativity Support Should be Explicit for Users}
Mode switches in our concept lead to different conversation flows (i.e. question-answer vs open dialogue), which can be confusing to users if not made clear. We improved on this in the second version of our prototype with special phrases that are revealed to users via suggestions in the UI. %

\highlightbox{1}{Introduce clear pathways in the (conversational) UI to switch between modes with different support roles for the creative writer.}

\subsection{Guided Prompts Support Round Characters yet Limit Impressions of a Conversation}
The guided prompts had a fixed question-answer structure. Sometimes the way users phrased the values did not fit this (e.g. regarding case grammar). Together, this limits the overall impression of having a conversation with the character in this mode. 
Nevertheless, immersion seemed less important for the prompts given that people saw the key value of this mode in \textit{discovering new attributes}. The suggested attribute \textit{values}, generated from a concept graph (see \cref{sec:implementation_guided}), gave similar benefits, including breaking out of author-typical tendencies.

\highlightbox{2}{Even conversation designs with very simple structure and clear boundaries can be inspiring for writers if targeted at ``one-word discoveries'', such as our character attributes here.}

As a critical reflection, do we need a chatbot for defining attributes? Our results suggest two benefits: (1) Defining attributes via chat integrates well with the chat UI for \textit{\modeOpen}. (2) Comments indicate positive user experience of chatting with a character instead of editing it as a table. Still, a table UI might help handling existing characters. Future work could explore further combinations of conversational UIs and other UIs in this context.

\subsection{Open Conversations are Inspiring yet Limited by Technical Capabilities}
Writers enjoyed the idea of conversing with their characters and the bot's open statements included inspiring elements. However, feedback showed that the generated text cannot achieve the impression of a realistic conversation consistently. From this we conclude that a main limitation are current text generation capabilities. Although the employed model took into account chat context and character information, its generated text did not always fit users' expectations. In the worst case, it might contradict specified attributes. %
Nevertheless, the improved prototype indeed achieved an enjoyable and interesting conversational impression for some writers, demonstrating a promising direction to support writers in this way. Finally, all writers pinned some statements by the bot as potentially interesting pieces for their character.

\highlightbox{3}{Use \textit{interaction possibilities} as a pragmatic response to (current) limitations in text generation capabilities for creative writing contexts: For instance, give writers explicit choice and control in the UI to handle AI generated text, with the added benefit of allowing them to explore more model output.}

Concretely, our improved UI implemented this by presenting generated text as suggestions, and showing multiple (i.e. three) such suggestions (\cref{fig:modes}b), which was positively received and allows for greater exploration of phrases. 
Finally, as text generation is a very active research area in NLP (e.g.~\cite{Brown2020language}), we expect that improved capabilities in the future could be used to generate better text, conditioned on character information, for this application context.

\subsection{Reflections on our Interaction Concept: Conversational AI as both Co-Creator and Creative Artifact}

We reflect on our concept through the lens of the framework for studying co-creative AI by \citet{Guzdial2019framework}: Therein, user and AI take turns to work on an ``artifact'' with two actions:
First, \textit{artifact actions} shape a created artifact (here: character). %
In our case, the user's actions define attributes and pin statements. The bot suggests attributes/values, and makes these statements. Thus, our bot's artifact actions always hand over initiative to the user. %
Second, \textit{other actions} do not directly affect the artifact. In our case, these include the attribute explanations  (e.g. \cref{fig:modes}a top). More subtly, those statements made by the bot that the user dismisses (e.g. choice 1 and 3 in \cref{fig:modes}b) do not affect the character -- and thus end up being non-artifact actions which still reveal the system's scope, in a role akin to ``What else?''-explanations~\cite{Lim2009}.
Overall, we thus realise a \textit{co-creative}~\cite{Guzdial2019framework} and \textit{mixed-initiative}~\cite{Horvitz1999} concept: User and AI take turns to shape the outcome. A unique novel aspect of our concept of \textit{\conceptBotShaping} %
here is that the chatbot-as-a-character is both a co-creator in the interaction as well as its outcome. Future work on character creation tools might build on this, for example, by also taking inspiration from work on chatbot conversation design (for a recent overview see~\cite{Choi2021}). %

\subsection{Reflections on Methodology \& Limitations}\label{sec:discussion_of_methods}

We employed an iterative user-centred design process with three empirical parts (online survey, two user studies). We reflect on these choices here. 
Our survey surfaced character attributes to complement those in the literature. Beyond that, it emphasised the varied approaches that writers use for character creation. This variation makes it challenging to derive very concrete design choices right away. We thus used an iterative approach, building on our core idea of a conversational UI. 
Our user studies revealed benefits, challenges, and opportunities. Study~1 informed UI/system changes which indeed improved the prototype, based on feedback and CSI scores. Moreover, most people used the tool in our studies to work on characters that are relevant to them, that is, for own writing projects. %
Despite this, task, size and duration of the study are limited: Future work could involve a larger sample, people who write as their main job, and longer use, to go beyond novelty and deeper into individual writing processes. %
The CSI~\cite{Cherry2014} quantified experiences as a score, and according to the obtained factor importances writers valued exploration and expression over immersion. %
We thus moved from immersion towards exploration in our revised prototype by presenting multiple bot utterances as suggestions. Since CSI and chat logging add little to the study duration, we recommend to use these methods to complement observations and interviews for such studies of creative writing tools.

%% file: sections/conclusion.tex
\section{Conclusion}
We have presented \textit{\system}, a chatbot that writers shape into a fictional character via conversation, a new concept we call \textit{\conceptBotShaping}.
It has two modes: First, prompts by the bot help writers define character attributes. Second, writers explore the character via open conversation that takes these attributes into account.
Two user studies showed \textit{\system}'s value as a support tool for writers: It was seen as particularly useful for ideation when creating new characters, and would benefit from improved (conditional) dialogue language models in the future, to achieve a better conversational impression.

More broadly, our work demonstrates a new concept of realising a co-creative AI system, considering conceptualisations in the literature~\cite{Guzdial2019framework}: Specifically, our AI is \textit{both} a co-creative actor (e.g. bot makes suggestions for the character) as well as the creative artifact acted upon (i.e. bot represents the character). 
This adds to the literature in two ways: 
First, recent discussions in HCI distinguish between AI as a tool and as emulation~\cite{Shneiderman2020aigrandgoals}. Here, we illustrate a new combination: \textit{\system} is a creative \textit{tool} for writers and an \textit{emulation} of the resulting character. 
Second, our interaction concept adds to the discussion around tools vs agents~\cite{Farooq2017chipanel, Shneiderman1997}: Using a conversational UI with two linked modes, our work provides a concrete point of reference for designing interactive AI to explore the conceptual space \textit{in between} tool-like (here: guided prompts) and partner-like (here: open conversation) interaction concepts~\cite{BeaudouinLafon2018chiworkshop}. 
We hope to stimulate further explorations in this direction by releasing our prototype and other material: \url{https://osf.io/dyatj/}